# Two types of proton-electron atoms in a vacuum and an extremely strong magnetic field

R. Fedaruk

*Institute of Physics, University of Szczecin, 70-451, Szczecin, Poland*

The Rutherford planetary model of a proton-electron atom is modified. Besides the Coulomb interaction of the point electron with the proton, its strong Coulomb interaction with the physical vacuum as well as the magnetic interaction between moving charges are taken into account. The vacuum interaction leads to the motion of the electron with the velocity of light $c$ in the circle with the radius being equal to the so-called classical electron radius $r_e$. Therefore, the velocity of the electron consists of two components: the velocity $\vec{\upsilon}$ of the mechanical motion and the velocity $\vec{c}$ of the photon-like motion. We postulate that $\vec{\upsilon} \perp \vec{c}$, and $\upsilon < c$. Hence, the electron inside the atom moves with the resulting faster-than-light velocity. The existence of two types of proton-electron atoms, the hydrogen atom and the neutron, is interpreted by the different motion and interaction of particles at large ($r \gg r_e$) and short ($r < r_e$) distances. In the first atom, the effect of photon-like motion is small, and the electron moves around the proton with the velocity $\upsilon \ll c$ in an orbit of the radius $r \gg r_e$. In the second atom, the photon-like motion is the determining factor, and the electron moves around the proton with the faster-than-light velocity in an orbit of the radius $r < r_e$. The calculated ground-state properties of the free hydrogen atom and the free neutron are in good agreement with the experimental data. The properties of these atoms in extremely strong magnetic fields ($B > 10^8$ T) are discussed.

## 1. Introduction

The hydrogen atom is the simplest two-body bound system consisting of a proton and an electron. The hypothesis of the second allotropic type of proton-electron atom, the strongly bound proton-electron atom called neutron, was suggested by Rutherford in 1920 [1]. The existence of the neutron was confirmed in 1932 by Chadwick, however the proton-electron model of the neutron was left because of its contradictions with quantum mechanics [2]. (a) According to quantum mechanics, the total angular momentum of the proton-electron neutron must be 0, while the neutron reveals as a particle with spin 1/2. (b) The binding energy of the proton-electron system must be negative, but the measured mass of the neutron is 0.782 MeV/c$^2$ larger than the sum of the masses of the proton and the electron. (c) Via the use of the spin magnetic moments of the proton and the electron it is impossible to obtain the neutron magnetic moment. (d) There is no possibility to reach the small neutron radius (of about $10^{-15}$ m) since the smallest radius of the proton-electron atom predicted by quantum mechanics and the Schrödinger equation is of $10^{-10}$ m. Although these contradictions are based on the theory describing only *weakly* bound atomic systems, they are generally accepted arguments against the possibility of existence of a *strongly* bound proton-electron atom.

Today, there are no experimental data related to the hydrogen atom which could give trouble to quantum electrodynamics (QED) [3]. At the same time, the precision of the measured proton and neutron properties, in particular their magnetic moments and the mass difference, is appreciably higher than the precision of their quantum chromodynamics (QCD) calculations [4,5]. Difficulties in describing nucleons as the strongly bound three-quark systems are usually related to computational problems.

Note that two alternative approaches, constructing nucleons either from the stable particles observed in the nucleon decay or from quarks nonexistent as free particles, confront with the same problem. That problem is the strong interaction of elementary constituents. Unlike the quark model, the strongly bound proton-electron model was definitely rejected. Nevertheless, some new attempts to improve of Rutherford's model of neutron were made [6–8]. In these attempts the problem of strong interaction between the electron and the proton was concealed by assuming the unusual structure of these particles. Recently, claims that so-called hydrino states exist have been discussed [9–12]. In such states the electron is strongly bound and is unusually close to the proton. It is obvious that, if strongly bound proton-electron states really exist as free atoms, their description could not be done in the frame of conventional approach.

On the other hand, quantum mechanics predicts strongly bound states of the hydrogen atom in strong magnetic fields ($B \gg 10^5$ T). Such states have been thoroughly studied theoretically using the Schrödinger and Dirac equations (see, e.g. Refs. [13–18]). In this case the Coulomb force is treated as a small perturbation compared to the magnetic force. Because of the strong magnetic confinement of the electron, the hydrogen atom attains a cylindrical structure elongated along the magnetic field and a much greater binding energy compared to the zero-field case. For example, in a field of about $1.2 \times 10^{10}$ T, the hydrogen atom becomes a long cylinder 200 times narrower than its normal diameter [17]. The highest magnetic fields (of about $10^2$ T) produced in laboratories are too small for the verification of theoretical results. The appropriate magnetic fields (up to $10^{13}$ T) can exist on the surfaces of neutron stars [18]. However, adequate experimental data confirmed the predicted properties of the hydrogen atom in such strong-field conditions were not found until now.

In the present paper, we discuss an opportunity of existence of strongly bound proton-electron states based on the modification of the Rutherford planetary model. Apart from the Coulomb interaction between the electron and the proton, an additional strong Coulomb interaction of charged particles with the physical vacuum as well as the magnetic interaction between moving charges should be taken into account. The atomic model presented predicts existence of two allotropic types of proton-electron bound states, the hydrogen atom and the neutron. The ground-state properties of the hydrogen atom and the neutron in a vacuum and in extremely strong magnetic fields are analyzed.

## 2. Allotropy of a proton-electron atom

In an attempt to search the origin of possible existence of two types of proton-electron atoms we accent the fact of duality of radiation and matter. It was experimentally proved that photons having energies equal or larger than $2m_ec^2$ can create electron-positron pairs or these pairs could annihilate into photons. Here, $m_e$ is the electron rest mass, and $c$ is the speed of light in vacuum. The duality of radiation and matter testifies that an electron and a photon are different manifestations of the same entity. Therefore, we postulate that, like a photon, a point electron possesses not only the rest energy $E = m_ec^2$, but also a corresponding momentum $|\vec{p}| = E/c = |m_e\vec{c}|$ resulting from the electron motion with the velocity of light (photon-like motion).

Let us assume that the photon-like motion of the electron is caused by its strong interaction with the environment (physical vacuum). This interaction can be expressed by means of the Coulomb force between elementary charges. The law of photon-like motion of the electron can be written in the following form:

$$\frac{m_ec^2}{r_e} = \frac{e^2}{4\pi\varepsilon_0 r_e^2}, \qquad (1)$$

where $e$ is the elementary charge, $\varepsilon_0$ is the electric constant. According to equation (1), the photon-like motion of the electron occurs in the circle with the radius being equal to the so-called classical electron radius $r_e = e^2/4\pi\varepsilon_0 m_ec^2 = 2.817940325(28) \times 10^{-15}$ m [19].

Due to its strong interaction with physical vacuum, an electron is never at rest and moves at any time with the velocity of light. Since the angular frequency of the motion is so high and its radius is so small, the photon-like motion of an electron cannot be directly measured. If the electron has no any additional motion, its average position does not change and it is observed as the resting electron.

It is obvious that in our model the radius $r_e$ of the circular motion of a point electron is not related to its structure. Historically, the term "classical electron radius" for the value $r_e$ was introduced in classical electrodynamics of an extended electron using the idea of its electromagnetic momentum and its electromagnetic mass [20–22].

The photon-like motion of an electron differs from the Zitterbewegung. According to the Dirac equation [23], a point electron, besides its slow motion, also possesses the oscillatory motion with the velocity of light (the Zitterbewegung [24]). This oscillatory motion occurs with the doubled Compton frequency $\omega_C = m_ec^2/\hbar$ and the amplitude of the order of the Compton wavelength $r_C = \hbar/m_ec$, where $\hbar$ is the reduced Planck constant.

It is important to point out that, in quantum theory, the radius of the Zitterbewegung is determined by the quantization of mechanical properties of an electron by means of the Planck constant. In our model, the radius $r_e$ of the photon-like motion of an electron is due to the quantization of an electrical charge and is much smaller than the Compton wavelength.

We now consider the motion of two point charged particles in the proton-electron atom. In general, the electromagnetic forces between two moving charged particles satisfy neither the condition that the forces are along the line joining them nor that these forces are equal in magnitude and oppositely directed [20–22]. There is a problem in the relation between Newton's third law and electromagnetism. However, the ground state of the proton-electron atom represents the extraordinary case of this relation. Due to the atom's stability, the atom is an isolate and closed system, and the emission of any electromagnetic radiation by the atom is absent. Therefore, we suppose that there is the full validity of Newton's third law for the interaction of moving charges inside the atom. As a result, at any instant, the forces of action and reaction between the electron and the proton are equal in magnitude and opposite in direction, and Newton's second law can be used to describe the motion of particles. Then, the centripetal force is determined by instantaneous values of velocities of the particles and is balanced by the Coulomb and Lorentz forces acting along the line joining the particles. Note that, owing to the exact validity of Newton's third law, the process of propagation of interaction between the particles inside the nonradiative atom can be excluded from the consideration. Consequently, Newtonian dynamics can be used in describing motions of the particles inside the atom and relativistic effects could be taken into account only in terms of the Lorentz force instead of special relativity.

In the atom, apart from the Coulomb and Lorentz interaction between the electron and the proton, the particles strongly interact with the physical vacuum, too. Therefore, the electron performs two types of motions and its instantaneous velocity consists of two components: the velocity $\vec{\upsilon}$ of the mechanical motion, and the velocity $\vec{c}$ of the photon-like motion. We postulate that these velocities are perpendicular to each other $\vec{v} \perp \vec{c}$, and $v < c$. Hence, inside an atom, the mechanical motion of the electron inevitably results in a faster-than-light velocity which does not exceed $\sqrt{2}c$. The same as the photon-like motion, the faster-than-light motion of the electron inside the atom cannot be directly measured, and it is necessary to recognize its manifestation in the world of conventional (much less than *c*) velocities. This motion leads to a modification of the classical planetary model and reveals in a different way in weakly and strongly bound atoms.

In a weakly bound (hydrogen) atom, the distance between the electron and the proton is large, $r \gg r_e$, and the quick motion of the electron with the velocity *c* in the circle of the radius $r_e$ is superimposed on its slow motion around the proton with the velocity $v \ll c$ in an orbit of the radius *r*. Therefore, the electron moves around the proton on a helical path. The velocity of the electron changes continuously its direction relative to the radius of the orbit. In result, only the instantaneous velocity of the electron is faster-than-light, but its orbital velocity is much less than *c*. Because of the averaging over the orbital period, the effect of photon-like motion on the properties of the hydrogen atom is small.

The faster-than-light velocity of the electron manifests quite differently when the distance between the electron and the proton is small, $r < r_e$. In such strongly bound (neutron) atom, the photon-like motion of the electron is the determining factor, and the electron circles the proton with a faster-than-light velocity. Since both velocities $\vec{v}$ and $\vec{c}$ are perpendicular to the orbital radius, its value does not change. A path of the electron lies on the surface of a sphere, but it is not flat.

Hence, the strong Coulomb electron-vacuum interaction gives an additional substantial contribution to the electromagnetic interaction between the proton and the electron, and results in the existence of two allotropic forms of proton-electron atoms.

### 3. Features of proton-electron atoms

The simultaneous existence of the mechanical and photon-like motions of the electron gives rise to difficulties in the precise analytical description of even the simplest, proton-electron atom. Therefore, only approximate analysis of two cases when $r > r_e$ and $r < r_e$ will be done. In the first case, the average effect of the photon-like motion of the electron is small and the slow motion of the electron around the proton with the velocity $v$ can be used as the zeroth approximation. In the second case, the photon-like motion of the electron must be taken into account from the outset.

Let us consider an electron and a proton are point particles. To simplify a description, the mass of the proton is assumed to be infinite. As it was argued above, for the atom in the nonradiative state, Newtonian dynamics can be used.

In the hydrogen atom, neglecting the photon-like motion, only the orbital motion of the electron with the velocity $v$ around the proton will be considered. This motion is balanced by the force of the electromagnetic interaction between the electrical charges. The electromagnetic force has two components. The first component, the Coulomb force $\vec{F}_C$, describes the electrostatic interaction. Since each moving charge produces a magnetic field, there exists also the second component, the Lorentz force $\vec{F}_L$, describing a magnetic interaction of the charges. The Lorentz force $\vec{F}_L$ is along the line connecting the charges and coincides in the direction with the Coulomb force $\vec{F}_C$. The magnetic field, created by an orbital motion of the proton in the position of the electron, is given by Biot-Savarte's law:

$$B = \frac{\mu_0}{4\pi} \frac{2\pi i}{r_H} = \frac{\mu_0}{4\pi} \frac{2\pi ef}{r_H} = \frac{1}{4\pi\varepsilon_0} \frac{e}{r_H^2} \frac{v}{c^2}, \quad (2)$$

where $\mu_0$ is the magnetic constant, $i = ef$, $f = v/(2\pi r_H)$, $r_H$ is the distance between the electron and the proton. Using formula (2), the Lorentz force can be written as

$$|\vec{F}_L| = |e\vec{v} \times \vec{B}| = \frac{e^2}{4\pi\varepsilon_0 r_H^2} \frac{v^2}{c^2}. \quad (3)$$

Newton's second law for the motion of the electron in the nonradiative hydrogen atom is

$$\frac{m_e v^2}{r_H} = F_C + F_L = \frac{e^2}{4\pi\varepsilon_0 r_H^2}\left(1 + \frac{v^2}{c^2}\right). \quad (4)$$

Using equation (4), the radius of the hydrogen atom can be written as

$$r_H = r_e \left(1 + \beta^2\right)/\beta^2, \quad (5)$$

where $r_e$ is the classical electron radius, $\beta = v/c$.

In the neutron, the distance between the electron and the proton is smaller than $r_e$. Due to the photon-like motion, the electron moves around the proton with the faster-than-light velocity

$$v_+ = \sqrt{c^2 + v^2} = c\sqrt{1 + \beta^2}. \quad (6)$$

The centripetal force, acting on the electron, is

$$F = \frac{m_e v_+^2}{r_+} = \frac{m_e c^2}{r_n} \sqrt{1 + \beta^2}, \quad (7)$$

where $r_+ = r_n \sqrt{1+\beta^2}$, $r_n$ is the distance between the proton and the electron in the neutron.

We assume that the Lorentz force, acting on the electron, is determined only by the velocity $\upsilon$ of the mechanical motion of the electron. However, the electromagnetic interaction between the electron and the proton is different at $r > r_e$, when the electron moves around the proton with the slower-than-light velocity, and at $r < r_e$, when the electron moves around the proton with the faster-than-light velocity. In the second case, the magnetic field created by the relative motion of the proton has to change its sign. Thus, the Lorentz force is described as usual by formula (3), but its direction is opposite to the Coulomb force. Then, the law of the motion of the electron in the neutron can be written as:

$$\frac{m_e c^2}{r_n} \sqrt{1+\beta^2} = F_C - F_L = \frac{e^2}{4\pi\varepsilon_0 r_n^2}\left(1-\beta^2\right). \quad (8)$$

From equation (8) we obtain the radius of the neutron:

$$r_n = r_e \left(1-\beta^2\right)/\sqrt{1+\beta^2}. \quad (9)$$

*Binding energy.* Forces, acting inside the atom, cannot be measured directly. Only the change of the energy of the atom or its total energy can be determined. The kinetic and rest energies of particles can be described by special relativity but the potential energy is not included in this theory. Thus, it is necessary to take into account the potential energy by a different way. Additionally, the Coulomb electron-vacuum interaction has to be included.

In the hydrogen atom, the kinetic energy of the electron consists of the energy of its mechanical motion, $m_e c^2/\sqrt{1-\beta^2} - m_e c^2$, and the energy of its photon-like motion, $m_e c^2$. The total kinetic energy of the electron is

$$E_1 = m_e c^2/\sqrt{1-\beta^2}. \quad (10)$$

The electromagnetic force given by the right part of equation (4) is along the line connecting the charges, and it is a central force. The velocity of the electron is perpendicular to the line connecting the charges and is constant. Therefore, using equation (4) the potential energy of the electron can be expressed as $U_1 = -e^2\left(1+\beta^2\right)/4\pi\varepsilon_0 r_H = -m_e \upsilon^2$. Similarly, the potential energy of the electron-vacuum interaction is equal to $-m_e c^2$. The total potential energy of the electron is

$$E_2 = -m_e \upsilon^2 - m_e c^2 = -m_e c^2\left(1+\beta^2\right). \quad (11)$$

The binding energy of the particles in the hydrogen atom is given by

$$\varphi_\infty = E_1 + E_2 = m_e c^2\left[1/\sqrt{1-\beta^2} - \left(1+\beta^2\right)\right]. \quad (12)$$

Since the electron-vacuum interaction has been neglected in the law of motion (Eq. 4), the energies of mechanical and photon-like motions of the electron in the hydrogen atom could be treated separately. In the neutron, the electron-vacuum interaction is the determining factor, and the mechanical and photon-like motions of the electron should be described simultaneously (Eq. 8). Consequently, the potential and kinetic energies could not be represented for the electron-proton and electron-vacuum interactions separately. Moreover, in order to obtain the measured energies, the energies of the faster-than-light electron should be normalized by the energies of the mechanical slower-than-light motion.

Using equation (8), the potential energy of the faster-than-light electron can be written as

$$U_2 = -\frac{e^2}{4\pi\varepsilon_0 r_n}\left(1-\beta^2\right) = -m_e c^2 \sqrt{1+\beta^2}. \quad (13)$$

This energy includes the electron-proton and electron-vacuum interactions. We have to modify the potential energy of the electron-vacuum interaction, $-m_e c^2$, taking into account the energy $U_2$ and normalizing this energy by the total kinetic energy $E_1$ of the slower-than-light motion given by equation (10). In this way, we obtain the measured part of the kinetic energy of the faster-than-light electron:

$$E_3 = -m_e c^2\left(U_2/E_1\right) = m_e c^2 \sqrt{1-\beta^4}. \quad (14)$$

We also have to modify the kinetic energy of the photon-like motion, $m_e c^2$, normalizing this energy by the total potential energy $E_2$ given by equation (11). Then the measured potential energy of the faster-than-light electron can be written as follows:

$$E_4 = m_e c^2 \left(m_e c^2/E_2\right) = -m_e c^2/\left(1+\beta^2\right). \quad (15)$$

The binding energy of the neutron is given by

$$\Phi_\infty = E_3 + E_4 = m_e c^2\left[\sqrt{1-\beta^4} - 1/\left(1+\beta^2\right)\right]. \quad (16)$$

The binding energy of the neutron does not include the kinetic energy of the mechanical motion of the electron. Such energy can be described separately and is

$$E_{m\infty} = m_e c^2\left(1/\sqrt{1-\beta^2} - 1\right). \quad (17)$$

It gives the significant contribution to the total energy of the neutron.

*Correction due to the finite proton mass.* The distance between the electron and the proton in the atom is

$$r = \xi_e + \xi_p, \quad (18)$$

where $\xi_e$ and $\xi_p$ are the distances that the electron and the proton are from the center of mass, respectively. According to the definition of the center of mass,

$$m_e \xi_e = m_p \xi_p. \quad (19)$$

From equations (18), (19) we obtain:

$$\xi_e = \frac{m_p}{m_p + m_e} r, \quad \xi_p = \frac{m_e}{m_p + m_e} r. \quad (20)$$

The linear velocities of the proton and the electron are:

$$\upsilon_p = \frac{m_e}{m_p} \upsilon_e = \frac{m_e}{m_p + m_e} \upsilon. \quad (21)$$

Equations (12), (16) and (17) corrected for the finite proton mass can be written as:

$$\varphi = \varphi_\infty - m_p c^2 \left[ 1/\sqrt{1-\beta_p^2} - (1+\beta_p^2) \right], \quad (22)$$

$$\Phi = \Phi_\infty - m_p c^2 \left[ \sqrt{1-\beta_p^4} - 1/(1+\beta_p^2) \right], \quad (23)$$

$$E_m = m_e c^2 \left( 1/\sqrt{1-\beta^2} - 1 \right) - m_p c^2 \left( 1/\sqrt{1-\beta_p^2} - 1 \right), \quad (24)$$

where $\beta_p = \upsilon_p / c$.

*Mass of atoms.* The rest energy of the hydrogen atom is defined as the sum of the rest energies of its constituent particles and the binding energy:

$$m_H c^2 = m_e c^2 + m_p c^2 + \varphi, \quad (25)$$

where $m_H$ is the mass of the atom. For bound states the binding energy is negative, and $m_H$ is smaller than the sum of the masses of the proton and the electron, giving rise to the mass defect.

The rest energy of the neutron consists of, in addition to the rest energies of the proton and the electron, the kinetic energy of the mechanical motion $E_m$ and the binding energy $\Phi$:

$$m_n c^2 = m_e c^2 + m_p c^2 + E_m + \Phi, \quad (26)$$

where $m_n$ is the neutron mass. Only the binding energy has negative values and ensures stable states of the neutron. The energy $E_m$ is positive, and much larger than the binding energy. Therefore, unlike the mass defect in the hydrogen atom, there is the mass "excess".

*Magnetic and angular moments of atoms.* The magnetic moment of the electron at its orbital circular motion is given by

$$\mu_\infty = -e\upsilon r^*/2, \quad (27)$$

where $r^* = r_H / \sqrt{1-\beta^2}$ for the hydrogen atom and $r^* = r_+ = r_n \sqrt{1+\beta^2}$ for the neutron.

Similarly, the orbital angular momentum of the electron is

$$L_\infty = m_e \upsilon r^*. \quad (28)$$

A correction due to the finite proton mass should be done both for the magnetic and angular moments. The correction for the angular momentum results in the replace of $m_e$ by the reduced mass. The masses of a proton and an electron have essentially different values, but the same signs. On the other hand, their electric charges have the same absolute values, but the different signs. Therefore, instantaneous positions of the center of mass and the center of charge do not coincide. In order to correct the position of the center of charge, it is necessary to change the position of the proton to the new position being symmetrical relative to the center of mass. Then, the distance between the electron and the proton is $r = \xi'_e - \xi'_p$, where $\xi'_e = m_p r / (m_p - m_e)$, $\xi'_p = m_e r / (m_p - m_e)$.

The magnetic and angular moments of the hydrogen atom ($L_H$, $\mu_H$) and the neutron ($L_n$, $\mu_n$) with the correction due to the finite proton mass can be written as

$$\mu_H = -\frac{ecr_e(1+\beta^2)}{2\beta\sqrt{1-\beta^2}} \frac{m_p + m_e}{m_p - m_e}, \quad (29)$$

$$\mu_n = -\frac{ecr_e \beta(1-\beta^2)}{2} \frac{m_p + m_e}{m_p - m_e}, \quad (30)$$

$$L_H = \frac{m_e c r_e (1+\beta^2)}{\beta\sqrt{1-\beta^2}} \frac{m_p}{m_p + m_e}, \quad (31)$$

$$L_n = m_e c r_e \beta(1-\beta^2) \frac{m_p}{m_p + m_e}. \quad (32)$$

## 4. Discussion

Our model describes the nonradiative states of the hydrogen atom and the neutron. The velocity of the electron specifies these states. The dependencies of radii of the atoms on the mechanical velocity $\upsilon$ of the electron are shown in Fig. 1. The radius $r_H$ of the hydrogen atom increases infinitely at $\upsilon \to 0$, but it tends to the value $2r_e$ at $\upsilon \to c$. On the other hand, the radius $r_n$ of the neutron at a small velocity is limited by the value $r_e$ and tends to zero at $\upsilon \to c$. Note that these atoms never have the same radius.

as well as the atomic radius and the Coulomb force, but the Lorenz force could never exceed the Coulomb force.

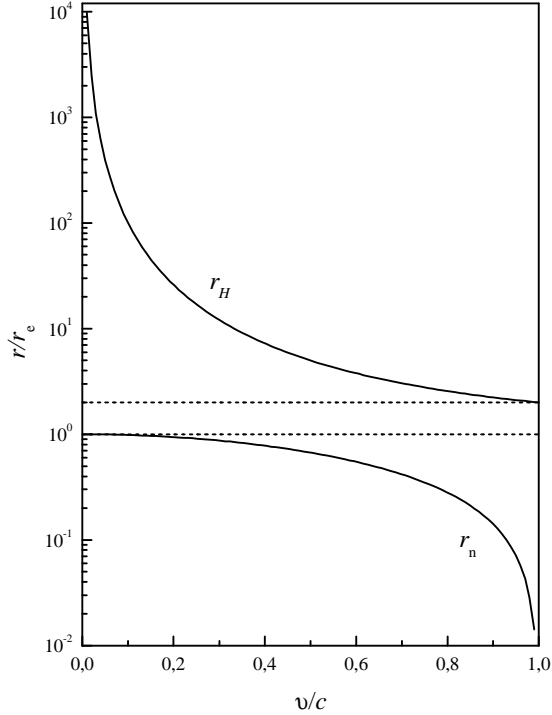

Fig. 1. Radii of the hydrogen atom ($r_H$) and the neutron ($r_n$) vs. the velocity of the electron given by formula (5) and (9).

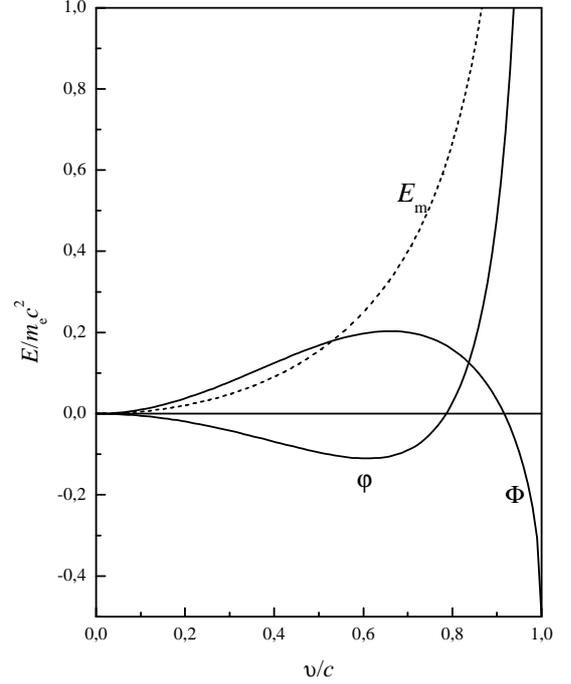

Fig. 2. Binding energies of the hydrogen atom ($\varphi$) and the neutron ($\Phi$) vs. the velocity of the electron. The dash line shows the kinetic energy of the mechanical motion of the electron in the neutron.

Fig. 2 shows the binding energies of the atoms as functions of the velocity $\upsilon$ of the electron. The dependencies demonstrate that the binding energy of the hydrogen atom is negative at $\upsilon \leq 0.8\ c$, whereas the binding energy of the neutron is negative at $\upsilon \geq 0.9\ c$. So, these atoms have stable states in a different range of velocity values. In the neutron, the kinetic energy of the mechanical motion $E_{m\infty}$ is much larger than the $|\Phi|$ at $\upsilon \geq 0.9\ c$, that results in the mass "excess".

The orbital velocity of the ground-state electron in the free atom is determined by the origin of this atom. An external magnetic field alters the orbital velocity of the electron in the free atom, thus transferring the atom into its new nonradiative state. The velocity of the electron increases or decreases depending on the orientation of the electron orbital magnetic moment relative to the magnetic field. In strong external magnetic fields, when the change of the orbital velocity of the electron cannot be treated as a perturbation, it is difficult to describe quantitatively the effect of external magnetic field. In this case the quantum model supposes that the Coulomb force can be treated as a small perturbation compared to the magnetic force [13–18]. As a result, the atomic structure is changed from spherical to cylindrical in extremely strong magnetic fields that are typical for neutron stars. In our model, a magnetic field does not change the symmetry of the atom. This field alters the orbital velocity of the electron

*Comparison with experimental data*. Measured values of the binding energy, the mass or the magnetic moment can be used to calculate the ground-state properties of the free hydrogen atom and the free neutron. As an illustration, the results of computations, which use measured values of the magnetic moments, are presented.

The measured ratio of the magnetic moment of the electron in the hydrogen atom $\mu_H^e$ to that of the free electron ($\mu_e = -1.00115965218111(74)\mu_B$ [19]) is equal to $1 - 17.709(13) \times 10^{-6}$ [3]. Consequently, $\mu_H^e = -1.001141943(13)\mu_B$, where $\mu_B$ is the Bohr magneton. Using this value and equation (29), we obtain $\beta_H = 0.00729755563(2)$. The radius of the hydrogen atom given by equation (5) is $r_H = 0.52917594(3) \times 10^{-10}$ m. According to equation (22), at $\beta = \beta_H$, the binding energy equals $-13.5985038(8)$ eV and is close to the experimental value $-13.59844$ eV [25]. The difference of the energies is comparable with the Lamb shift $8172840(22)$ kHz ($3.38 \times 10^{-5}$ eV) [26].

Using the experimental value of the neutron magnetic moment [19] $\mu_n = -1.9130427(5)\ \mu_N$, ($\mu_N$ is the nuclear magneton), for the neutron, we found the following values: $\beta_n = 0.91914955(3)$, $r_n = 0.321911718(8) \times 10^{-15}$ m, $\Phi_n = -3826.22(3)$ eV, $(m_n - m_p)c^2 = 1.2933083(3)$ MeV. The experimental neutron-proton mass difference is $(m_n - m_p)c^2 = 1.2933317(5)$ MeV [19]. The calculated value $r_n^2$

= 0.103627 fm$^2$ correlates with the root-mean-square charge radius of a neutron $\langle r_n^2 \rangle$ = – 0.1161(22) fm$^2$ obtained from elastic electron-deuteron scattering using QCD [19]. The value $r_n / \sqrt{1-\beta_n^2}$ = 0.8172379(3) fm correlates with the magnetic root-mean-square radius of the neutron 0.873(15) fm [27].

The calculated physical quantities characterizing the hydrogen atom and the neutron are in good agreement with the experimental data. However, in more precise description, some fine effects should be taken into consideration. In our model, only the static electron-vacuum interaction was taken into account. Besides, the electron interacts with vacuum fluctuations. This interaction with virtual photons and electron-positron pairs has a stochastic nature and is described by QED. The static and fluctuating vacuum interactions exist simultaneously, result in completely different physical effects and require two different (deterministic or probabilistic) descriptions. In particular, vacuum fluctuations cause "the trembling" of the electron resulting in the Lamb shift of atomic levels. We estimate that in the hydrogen atom the effect of vacuum fluctuations is stronger by factor of about 10$^3$ than that of the photon-like motion of the electron, and that could be the main reason of the small discrepancy between our calculations and the experimental results.

On the other hand, our results disclose the substantial difference between the inner (non-observed, hidden) and observed mechanical properties of the proton-electron atoms. According to the planetary model, if the mass and the electric charge of the electron are localized in the same point, its orbital angular momenta and magnetic moments have similar dependencies on its orbital motion. The calculated orbital magnetic moments of the electron in the hydrogen atom and in the neutron are close to their experimental values. But, as it follows from equations (31) and (32), the respective orbital angular momenta are $L_H \approx 0.9995\,\hbar$ and $L_n \approx 1.0413 \times 10^{-3}\,\hbar$ and do not coincide with the measured value of $\hbar/2$.

One possibility to reconcile these features is the assumption that the observed mechanical and magnetic properties relate to two different entities. We assume that the measured gyromagnetic ratio of the bound electron is the ratio of its orbital magnetic moment to the angular momentum $\hbar/2$ resulting from the process of resonant absorption or emission of photons. So, the orbital magnetic moment of the electron is directly measured in magnetic-resonance experiments. Quite the contrary, the finite orbital angular momentum, though really exists, does not change the energy of the magnetic moment in the magnetic field and is non-observable. Unlike quantum mechanics, in our concept the angular momentum $\hbar/2$ relates to the process of the resonant absorption of electromagnetic radiation rather than to the own properties of the electron. Therefore, it does not depend on the orbital motion and it is the same for the free electron or the electron bound in any atoms. Note that the unjustified attribution of the angular momentum $\hbar/2$ to the own properties of the electron did not allow until now to propose the consistent physical model of the spin though numerous attempts were made. Moreover, as it is known, projections of spin angular momenta of all particles are multiple of $\hbar/2$ and do not depend on any of their properties (mass, charge, structure, magnetic moment, etc.). At the same time, masses and magnetic moments have no multiplicity, i.e. their quanta are absent. Such strong difference between the mechanical and magnetic properties has no physical interpretation.

*Electric and magnetic fields inside atoms.* We shall now discuss the relation between atomic properties and extremely strong electromagnetisms. We found that, in hydrogen atom, the orbital motion of the electron is relatively slow (about 137 times slower than the light), the magnetic field $B_H$ is of 12.5 T and its effect is small in comparison with the effect of the electrostatic field ($E_H$ = 5.14 × 10$^{11}$ V/m). The Lorentz force $F_L$ is about of 0.53 × 10$^{-4}$ $F_C$. In a neutron, the electron moves in extremely strong electric ($E_n$ = 1.39 × 10$^{22}$ V/m) and magnetic ($B_n$ = 4.26 × 10$^{13}$ T) fields exceeding appreciably the QED critical values ($B_{cr}$ = 4.4 × 10$^9$ T, $E_{cr}$ = 1.3 × 10$^{18}$ V/m). In this case the influence of the magnetic field is comparable to the electrostatic interaction; $F_L$ is about of 0.84 $F_C$.

The faster-than-light motion of the electron and the extremely strong electric and magnetic fields inside the neutron result in its unusual properties and give rise to the fundamental contradictions between the proton-electron model of the neutron and quantum mechanics. These contradictions indicate restrictions of quantum mechanics beyond the description of *free* particles and *weakly* bound atomic systems. This theory has problems in describing *strongly* bound systems and cannot give arguments against the proton-electron model of the neutron.

As it is easy to see, physics of extremely strong electromagnetisms, existing according to our model at short distances, $r < r_e$, merges with physics of strong interactions given in a different way by QCD. The force of interaction between the proton and the electron into the neutron (Eq. (8)) represents basic properties attributed to the interaction between quarks in QCD and called the confinement and the asymptotic freedom. When $\upsilon \to 0$, the Lorentz force vanishes, only the Coulomb force acts and the radius $r_n$ of the orbit of the electron tends to its maximum value $r_e$. When $\upsilon \to c$, the Lorentz force compensates the Coulomb force, and the radius $r_n$ tends to zero. Most likely, even at very short distances, the objective reality is not described by one model only, but it is necessary to reconcile the deterministic and probabilistic aspects.

## 5. Conclusions

In the planetary model of the proton-electron atom, we take into account, besides the Coulomb interaction of the point electron with the proton, its strong Coulomb interaction with the physical vacuum as well as the magnetic interaction between moving charges. These modifications allow us to construct its two different bound states: the hydrogen atom and the neutron. Our model enables to analyze from the new point of the simplest bound states in atomic and nuclear physics. Such approach might be useful for better understanding static properties of nucleons and low-energy hadron physics where QCD is confronted with difficulties.

I am grateful to K. Czerski and M.P. Dąbrowski for reading the manuscript and useful comments and discussions.